\documentclass[trackchanges]{aastex701}  %%linenumbers,twocolumn

\usepackage{bm}
\usepackage{graphicx}
\usepackage{CJKutf8}    %% pdflatex

\begin{document}
\begin{CJK*}{UTF8}{gbsn}

\title{Simulations of Particle Acceleration by an Interplanetary Shock Observed at 0.07 au and 0.7 au}

\author[orcid=0000-0001-7617-8268,sname='China']{F.-J. Kong (孔凡婧)}
\affiliation{School of Electronic Engineering, North China 
University of Water Resources and Electric Power, Zhengzhou, 450046; kongfanjing@ncwu.edu.cn}
\email[show]{kongfanjing@ncwu.edu.cn}  

\author[orcid=0000-0002-3437-3716,sname='China']{G. Qin (秦刚)} 
\affiliation{School of Science, Harbin Institute of Technology,
Shenzhen, 518055; qingang@hit.edu.cn, ywangsz@hit.edu.cn}
\affiliation{Shenzhen Key Laboratory of Numerical Prediction for
Space Storm, Harbin Institute of Technology, Shenzhen, 518055}
\email[show]{qingang@hit.edu.cn}

\author[orcid=0000-0002-4581-9242,sname=China]{Y. Wang (汪洋)}
\affiliation{School of Science, Harbin Institute of Technology,
Shenzhen, 518055; qingang@hit.edu.cn, ywangsz@hit.edu.cn}
\affiliation{Shenzhen Key Laboratory of Numerical Prediction for
Space Storm, Harbin Institute of Technology, Shenzhen, 518055}
\email[show]{ywangsz@hit.edu.cn}

\begin{abstract}

Parker Solar Probe (PSP) observed a coronal mass ejection (CME)-driven shock
at 0.07 au on 2022 September 5, and the same CME-driven interplanetary shock
was subsequently observed by Solar Orbiter (SolO) at 0.7 au on 2022 September 6. 
Such a large difference in heliocentric distance between the two shock events enables 
our comparative study of particle acceleration in near-Sun and far-Sun regions. 
Using a backward-in-time test-particle approach in a slab-2D turbulent magnetic 
field model, we perform particle acceleration simulations for the 0.07 au PSP shock 
and 0.7 au SolO shock, covering the energy ranges measured by PSP/EPI-Lo 
(0.0697--8.06 MeV) and SolO/EPD (0.0495--5.83 MeV). With an initial $\kappa=3$ 
kappa distribution, the particle fluxes at the shock passage are successfully 
reproduced using effective local shock angles of $70^\circ$ for the PSP shock and
$80^\circ$ for the SolO shock. Both the observed and simulated energy spectra exhibit
a double power-law form. The 0.07 au PSP spectrum softens above $\sim$0.75 MeV, 
while the 0.7 au SolO spectrum hardens slightly above $\sim$0.51 MeV, ‌revealing 
distinct near-Sun and far-Sun acceleration environments. Compared with the 
observations, the simulations yield harder energy spectra with higher break energies
of $\sim$2 MeV at 0.07 au and $\sim$3 MeV at 0.7 au, indicating efficient particle
acceleration under the initial $\kappa=3$ distribution. Both the observed and simulated
spectral indices are harder than the standard diffusive shock acceleration (DSA) 
prediction, likely arising from non-local acceleration at the large-angle 
quasi-perpendicular shocks, efficient injection of pre-accelerated suprathermal 
particles, and the finite shock lifetime effect.

\end{abstract}


\section{Introduction} 

Flux enhancements of energetic particles in interplanetary space 
are usually accompanied by shock waves, which are important acceleration sites
of particles. Solar energetic particle events, as an important part
of space weather, consist of impulsive events originating in solar flares 
and gradual events associated with shocks driven by coronal mass ejections (CMEs). 
CME-driven shocks experience diverse interplanetary environments 
during their outward propagation, which contributes to the
complex evolution of particle fluxes. Typical related phenomena include
the 90$^\circ$ pitch angle enhancements of suprathermal electrons 
caused by shock drift acceleration at quasi-perpendicular shocks \citep{YangEA2018,KongAQin2020}, 
dropout in energetic-ion ﬂuxes likely due to the interaction 
of the shock with a magnetic structure \citep{GiacaloneEA2021},
and inverse velocity dispersion feature for energetic protons 
primarily caused by diffusive shock acceleration process \citep{LiEA2025}.

The widely accepted diffusive shock acceleration theory \citep[DSA;][]{AxfordEA1977,Krymsky1977,Bell1978,Blandford1978} includes first-order Fermi acceleration and shock drift acceleration mechanisms. 
The shock geometry directly influences the acceleration process: it is 
dominated by first-order Fermi acceleration at quasi-parallel shocks 
and shock drift acceleration at quasi-perpendicular shocks \citep{Decker1986,Ohira2016}.
The turbulence level also affects the acceleration mechanism in different 
shock geometries \citep{QinEA2018,KongAQin2020}. There are two other acceleration
mechanisms, i.e., stochastic acceleration and shock surfing acceleration. 
Stemming from the turbulence downstream of the shock, stochastic acceleration 
can explain the generation of energetic particles in solar flares \citep{Petrosian2016}. 
Particles will undergo shock surfing acceleration when a finite cross-shock 
potential exists within the shock transition layer. Using the focused 
transport equation, \citet{ZuoEA2013} modeled the acceleration of 
low-energy ions at parallel shocks, and found that the cross-shock potential 
can noticeably enhance the efficiency of particle injection.

The particle injection problem is an unresolved puzzle in shock acceleration research.
For quasi-parallel shocks, a fraction of injected particles can be 
reflected upstream as they traverse the shock transition layer. 
There is, however, the injection problem for quasi-perpendicular shocks, 
because particles tied to magnetic field lines will directly enter 
the downstream of the shock via solar wind convection. 
\citet{Giacalone2005} found that particles in the large-scale magnetic fluctuations 
can encounter the shock front multiple times by following meandering field lines, 
and therefore those particles can be diffusively accelerated to high energies. 
Some low-energy particles near the shock front may undergo pre-acceleration 
through drift acceleration caused by magnetic gradient or shock surfing 
acceleration due to cross-shock electric potential. 
Therefore, factors including the shock geometry, magnetic turbulence level 
and cross-shock electric potential can significantly influence the injection
of low-energy particles into diffusive shock acceleration process.

The origin of particles energized by interplanetary shocks remains a critical issue.
These particles can originate from thermal populations in the solar wind, 
pre-existing superthermal ions, or a mixture of both \citep{Desai2016}.
The particle intensity observed downstream of the shock is probably 
attributed to the continuous acceleration of solar wind thermal particles. 
Meanwhile, the shock is ‌often embedded in an interplanetary environment 
rich in superthermal particles when it propagates outward from the near-Sun region. \citet{Kahler2019} confirm that suprathermal intensities measured near 1 au 
serve as a reliable observational indicator for forecasting SEP event peak intensities.
Under the assumption of a valid diffusion approximation, particles can be 
injected into the DSA process with an initial Maxwellian distribution in 
the solar wind, or alternatively a kappa distribution with a suprathermal tail 
\citep{Neergaard-Parker2012,Neergaard-ParkerEA2014}. 
The preexisting superthermal particles can be observed far upstream of parallel 
shocks and exhibit intensity peaks prior to the arrival of the shock front. 
However, for perpendicular shocks, the intensity enhancements of superthermal 
particles are only detected immediately upstream of the shock front \citep{LarioEA2019}. 
It is still not conclusive whether superthermal particles can be considered
the seed populations for the acceleration of particles at shocks. 
Recent explorations such as Parker Solar Probe and Solar Orbiter 
spacecrafts within 1 au provide observational evidences for understanding 
near-Sun particle intensity distributions during shock acceleration.

Particle acceleration mechanisms at collisionless shocks are strongly modulated
by the shock geometry. However, in the electron acceleration study of two
Wind-observed CME-driven interplanetary shocks, one quasi-perpendicular and the other 
quasi-parallel, presented by \citet{YangEA2018}, the ratio of downstream to 
upstream intensities peaks near the 90$^\circ$ pitch angle, and the downstream
energy spectral index is much larger than the theoretical index predicted 
by DSA. Our previous simulations in \citet{KongAQin2020} confirmed the dominant 
role of shock drift acceleration for the quasi-perpendicular shock event reported in \citet{YangEA2018}.
In previous studies, the understanding of the shock acceleration process
within 1 au near the Sun was limited due to the lack of in-situ observational data. 
The launches of Parker Solar Probe (PSP) in 2018 and Solar Orbiter (SolO) in 2020
provide a unique opportunity to investigate the near-Sun shock acceleration process. 
In this paper, we perform test particle simulations to investigate particle 
acceleration at the interplanetary shock observed by PSP at $\sim$0.07 au 
on 2022 September 5 and later by SolO at $\sim$0.7 au on 2022 September 6 \citep{TrottaEA2024}. 
Section \ref{sec2} introduces the in-situ data associated with these two
shock events. The numerical model is described in Section \ref{sec3}. 
Section \ref{sec4} presents a detailed comparison of energy spectra between the 
simulations and observations for the interplanetary shock at the two distinct heliocentric distances. 
In this section, we further investigate how the average energy gain 
and cumulative residence time within one gyroradius of the shock front 
depend on the shock angle, shock compression ratio, and upstream Alfvén Mach number, 
for 0.1-MeV, 1-MeV, and 10-MeV particles. 
Finally, the summary and discussion are given in Section \ref{sec5}.

\section{Observations} \label{sec2}
In this section, we report magnetic field, plasma, and particle-intensity data 
for the interplanetary shock observed by PSP on 2022 September 5 at 0.07 au and by SolO 
on 2022 September 6 at 0.7 au. At PSP, we use 1-minute resolution magnetic field 
data from the FIELDS instrument \citep{BaleEA2016}, and proton density, flow velocity, 
and temperature data from Solar Probe ANalyzer-Ions \citep[SPAN-I;][]{LiviEA2022} 
on the Solar Wind Electrons Alphas and Protons (SWEAP) investigation suite \citep{KasperEA2016}.
At SolO, 1-minute resolution magnetic field data are measured by flux-gate
magnetometer \citep[MAG;][]{HorburyEA2020}, and density, velocity, and temperature are 
measured by the Solar Wind Analyser suite \citep[SWA;][]{OwenEA2020}. The average 
flow velocity, magnetic field, density, and temperature values are obtained from 
the observations within 10 minutes before the arrival of the shock, as listed in 
rows 4 and 6--8 of Table \ref{tab:shockpara}. The shock angle, compression ratio, 
shock speed, and Alfvén Mach number listed in rows 1--3 and 5 of Table \ref{tab:shockpara} 
are taken from the work of \cite{TrottaEA2024}.

The proton fluxes on the 2022 September 5 shock event in the energy range of 
0.07--8 MeV are obtained from the Energetic Particle Instrument-Low Energy 
(EPI-Lo) instrument \citep{HillEA2017} on the Integrated Science Investigation 
of the Sun instrument suite \citep[IS$\Theta$IS;][]{McComasEA2016} onboard PSP.
The EPI-Lo instrument features a wide sky coverage achieved via 80 apertures.
In this study, we utilize flux data from EPI-Lo Channel P, which provides the 
highest available cadence and is exclusively optimized for proton measurements. 
The ion fluxes on the 2022 September 6 shock event in the energy range of 0.05--5.8 MeV 
are measured by the EPT sensor on the Energetic Particle Detector suite \citep[EPD;][]{Rodriguez-PachecoEA2020} 
onboard SolO. The EPD-EPT instrument is designed to sample four distinct fields 
of view, namely the Sunward, Anti-Sunward, North and South directions. 
All of the above observational data are available on the website
of Coordinated Data Analysis Web (CDAWeb) (\url{https://cdaweb.gsfc.nasa.gov/}).

\begin{deluxetable*}{lll}
\tablecaption{Shock Parameters\label{tab:shockpara}}
\tablehead{
\colhead{Parameters} & \colhead{Parker Solar Probe} & \colhead{Solar Orbiter}}
\startdata
$\theta_{\text{Bn}}$    & 53 & 51 \\
$s$  & 2.3 & 1.9 \\
$V_{\text{sh}}$ (km s$^{-1}$) & 1520 & 942 \\
$U_{\text{sw}}$ (km s$^{-1}$) & 340  & 480 \\
$M_{\text A1}$    & 3.9  & 3.8 \\
$B_0$ (nT)    & 608 & 15 \\
$n_{\text p}$ (cm$^{-3}$) & 1216 & 3.9 \\
$T_{\text p}$ (eV)        & 66 & 60 \\
$\lambda_{\text{slab}}$ (m) & $2.1\times10^8$ & $2.1\times10^9$
\enddata
\end{deluxetable*}

\begin{figure*}[!htb]           
 \centering
\includegraphics[scale=0.6]{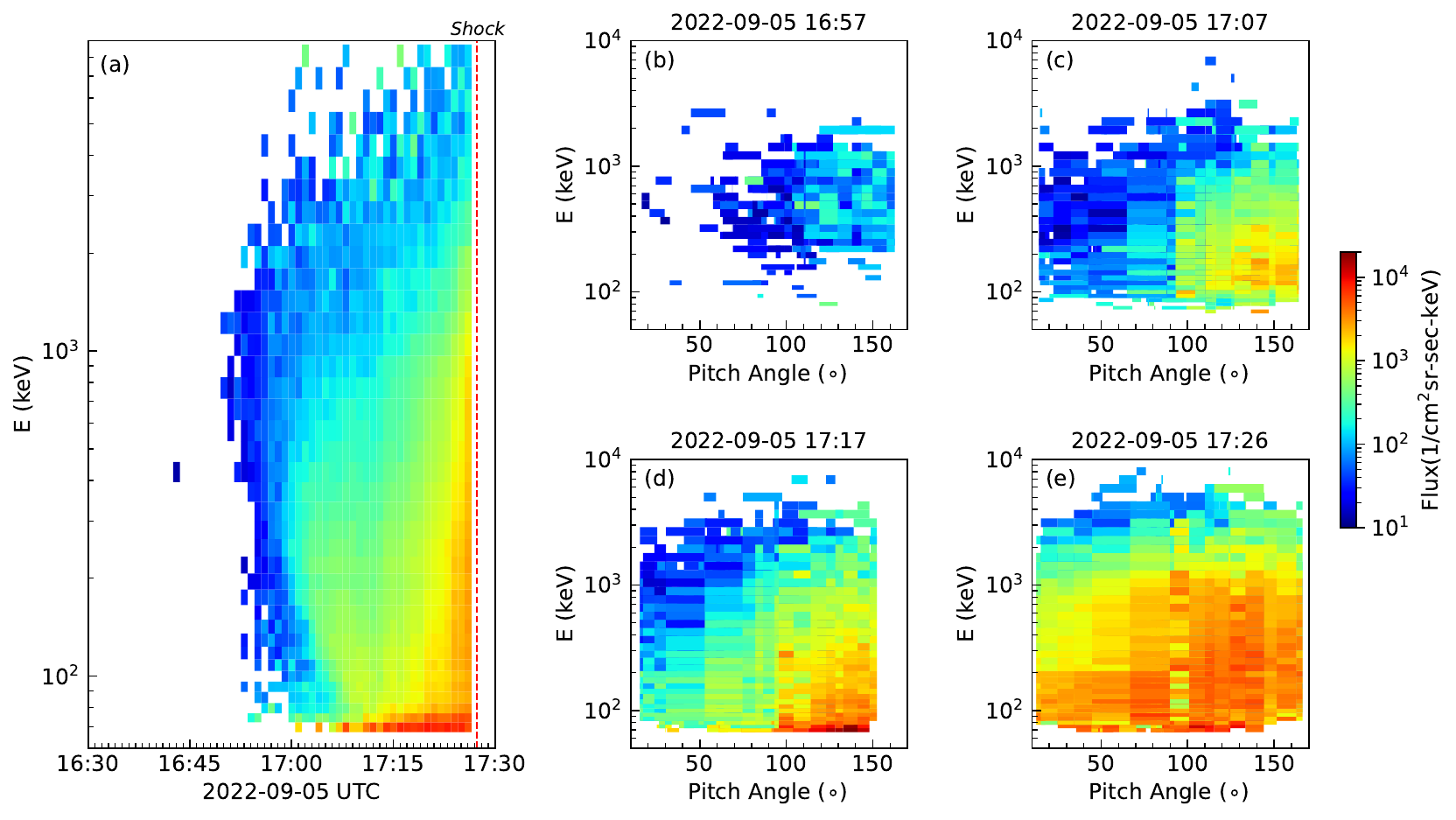}
\caption{(a): Energy spectrogram of average proton fluxes in the look directions 
of 80 apertures measured by EPI-Lo instrument in the Channel P on PSP/IS$\Theta$IS
before the shock arrival on 2022 September 5. The time of shock arrival is marked
by the red dashed line.
(b)--(e): Energy spectrograms of proton fluxes in all pitch angles obtained from
PSP/IS$\Theta$IS/EPI-Lo instrument at the time 16:57(b), 17:07(c), 
17:17(d), and 17:26(e) on 2022 September 5.
\label{fig:fluxEtpitch}}
\end{figure*}

Figure \ref{fig:fluxEtpitch}(a) shows the energy spectrogram of proton fluxes 
measured by the EPI-Lo instrument on PSP/IS$\Theta$IS before the arrival of the 
shock on 2022 September 5. At energies below $\sim$1 MeV, higher-energy protons 
arrived earlier, whereas at energies above $\sim$1 MeV, lower-energy protons 
exhibited earlier arrival. This shows an inverse velocity dispersion above 
$\sim$1 MeV, as discussed in the recent studies \citep{CohenEA2024,KouloumvakosEA2025}. 
As CME-driven shocks propagate outward into interplanetary space,
they can continuously trap the surrounding charged particles and 
efficiently energize them through DSA process. 
Low-energy particles below 1 MeV can reach their target energy after only
a few shock crossings, enabling rapid escape from the shock front and 
leading to an early release time. Since these low-energy particles are released 
nearly simultaneously, their arrival times follow a normal dispersion pattern. 
In contrast, particles above 1 MeV require continuous acceleration by the shock, 
so they escape the acceleration region much later. This contributes to a 
systematic delay with increasing particle energy, producing the characteristic 
inverse dispersion signature. 
Shown in Figure \ref{fig:fluxEtpitch}(b)--(e) are the energy spectrograms of 
proton fluxes at all pitch angles at four discrete moments: 16:57, 17:07, 
17:17 and 17:26, prior to the shock arrival at 17:27:19. At 16:57 (Figure 
\ref{fig:fluxEtpitch}(b)), the proton fluxes in all energy channels at each 
pitch angle direction are relatively low, with most values below $\sim$$10^2$ cm$^{-2}$sr$^{-1}$s$^{-1}$keV$^{-1}$. 
At 17:07 and 17:17 (Figure \ref{fig:fluxEtpitch}(c)--(d)), 
the fluxes at pitch angles above 100$^\circ$ show a pronounced increase. 
Just ahead of the shock arrival at 17:26 (Figure \ref{fig:fluxEtpitch}(e)), 
the fluxes at all pitch angles show a significant enhancement with values 
up to $\sim$10$^4$ cm$^{-2}$sr$^{-1}$s$^{-1}$keV$^{-1}$.

\begin{figure*}[]           
 \centering
\includegraphics[scale=0.6]{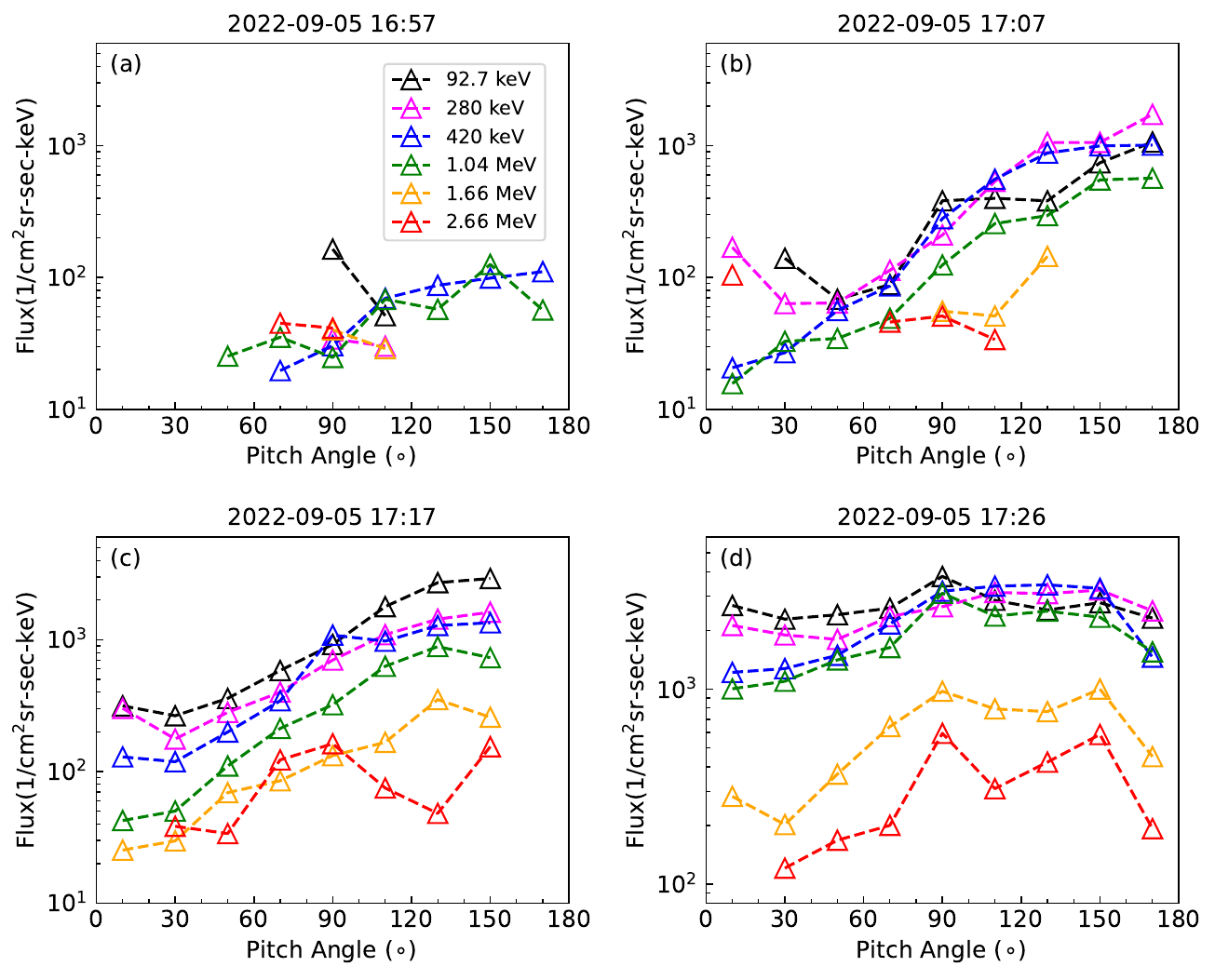}
\caption{Pitch angle distributions in several energy channels at 
16:57(b), 17:07(c), 17:17(d), and 17:26(e) on 2022 September 5.
Note that in each energy channel, the average particle flux is calculated 
in $20^\circ$ pitch angle bins using the observational data from the 
PSP/IS$\Theta$IS/EPI-Lo instrument.
\label{fig:fluxpitch}}
\end{figure*}

Figure \ref{fig:fluxpitch} presents the pitch angle distributions at several 
energy channels (i.e., the central energy of each energy bin) 
at the times illustrated in panels (b)--(e) of Figure \ref{fig:fluxEtpitch}.
The average particle flux is calculated in $20^\circ$ pitch angle
bins using measurements from the EPI-Lo instrument onboard PSP.
At 16:57, for 92.7 keV, 1.66 MeV, and 2.66 MeV particles the fluxes 
are confined exclusively to the pitch angle range of $60^\circ$--$120^\circ$.
At 17:07, for 92.7 keV--1.04 MeV energy range, the fluxes extend 
over a wide pitch angle interval, exhibiting an obvious particle
beam aligned along the 180$^\circ$ pitch-angle direction.
By 17:17, the particle beam persists in all energy channels below 1.66 MeV.
Immediately prior to the shock arrival at 17:26, the pitch-angle anisotropy weakens
substantially, and the flux near the $90^\circ$ pitch angle shows a slight
enhancement. A distinct double-peak signature emerges in the 1.66 MeV 
and 2.66 MeV energy channels. 
During the acceleration process, particles are continuously scattered by
magnetic turbulence, so their pitch-angle distribution tends to be isotropic. 
However, since shock drift acceleration mechanism is more effective, 
particle fluxes show a slight enhancement at the $90^\circ$ pitch angle
for 92.7 keV particles. For particles above 280 keV, higher fluxes are observed
in the $80^\circ$--$160^\circ$ pitch angle range, which is attributed to the 
field-aligned particle beams prior to the shock arrival.

\begin{figure*}[ht!]
 \centering
\includegraphics[scale=0.6]{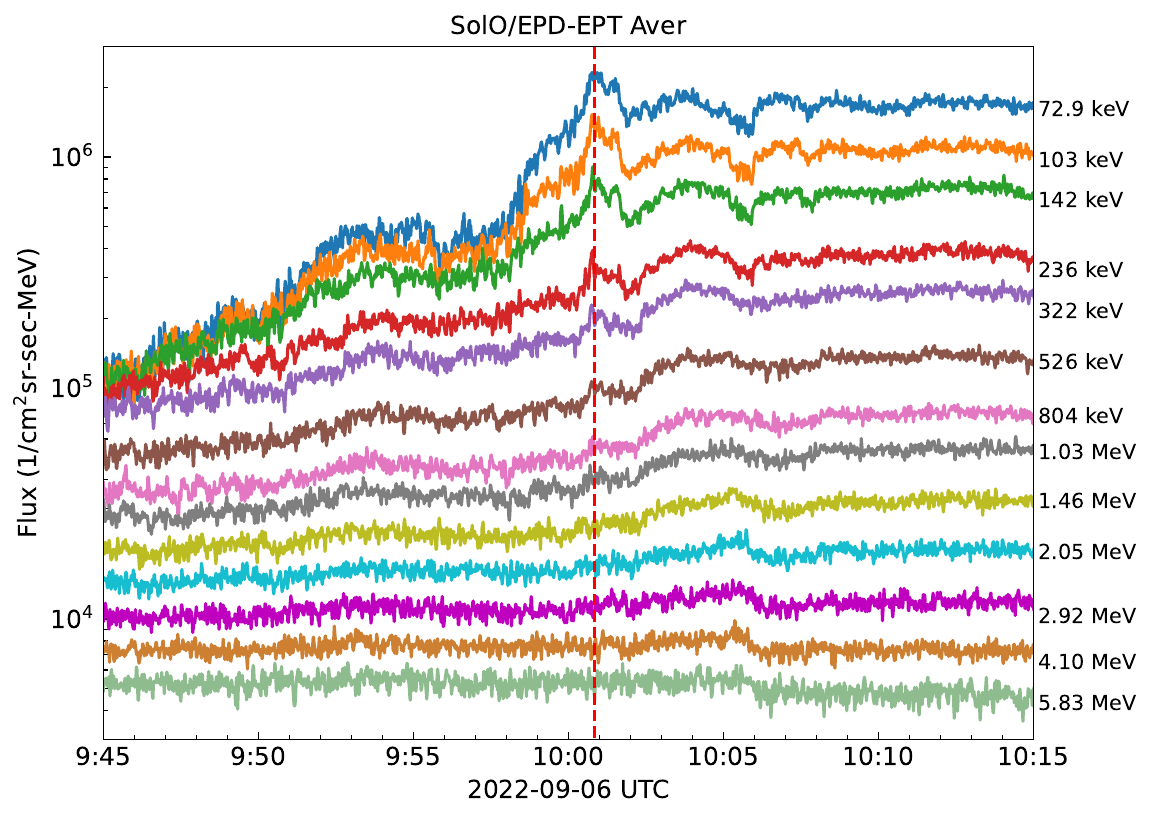}
\caption{Variations of ion (mostly proton) fluxes with time in several energy 
channels marked on the right measured by SolO/EPD/EPT instrument on 2022 September 6. 
Note that the flux is averaged over Sunward, Anti-Sunward, North and South directions. 
The vertical red dashed line corresponds to the shock crossing of SolO.
\label{fig:fluxtsolo}}
\end{figure*}

Figure \ref{fig:fluxtsolo} shows the temporal variations of ion fluxes 
in the energy range of $\sim$70 keV to 5.8 MeV, obtained from EPT 
measurements on the EPD instrument onboard SolO. 
The ion flux at energies below 526 keV exhibits a sharp peak exactly 
at 10:00:51 UT on 2022 September 6. This is a typical signature of
low-energy particle acceleration and rapid local flux enhancement
at the shock passage by SolO. For particles with energies above 804 keV, 
especially those above 1.03 MeV, no distinct flux peak is observed at 
the time of shock arrival. This feature is closely related to the shock 
acceleration ability, indicating that the maximum energy threshold in 
this shock event is lower than 1 MeV. 
As listed in Table \ref{tab:shockpara}, the shock compression ratio 
is merely 1.9, a value that is likely inadequate to facilitate efficient 
shock acceleration of particles above 1 MeV.

\section{Numerical Model} \label{sec3}
\begin{figure*}[ht!]
 \centering
\includegraphics[scale=0.6]{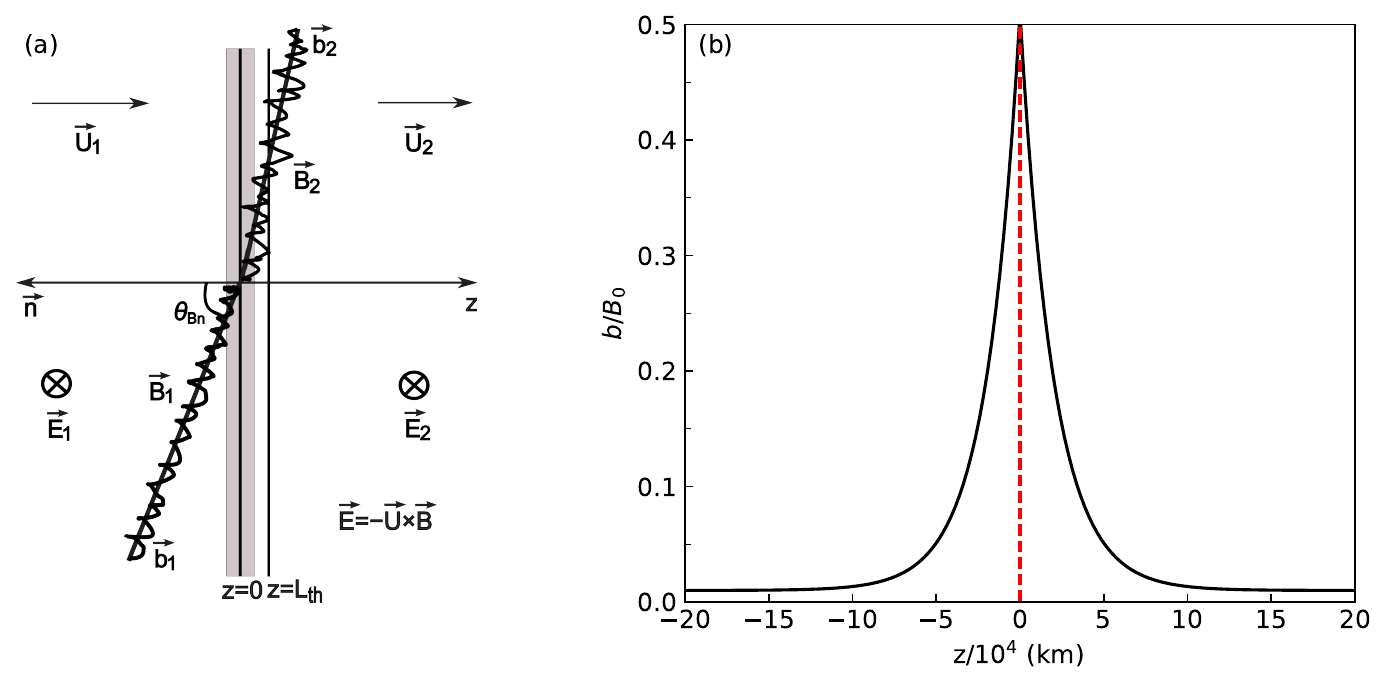}
\caption{(a): Shock geometry in the simulations. The gray shadow
indicates the shock transition. Particles are released at the location 
$z=L_{\text{th}}$, where $L_{\text{th}}$ is the thickness of the shock transition.
(b): Variations of the amplitude of magnetic field fluctuations $b/B_0$ with position $z$,
where $b$ is the amplitude of magnetic field fluctuations and $B_0$ is the background magnetic field.
Note that $z<0$ and $z>0$ correspond to upstream and downstream of the shock, respectively. 
\label{fig:shockdbB0}}
\end{figure*}

We study the acceleration of protons on the 2022 September 5 shock event 
observed by PSP at 0.07 au, and 2022 September 6 shock event observed 
by SolO at 0.7 au. The trajectories of particles are calculated by solving 
the equation of motion in a pregenerated static electromagnetic field
\begin{equation}
 \frac{d\bm p(\bm r,t)}{dt}=q[\bm{E}(\bm{r})+\bm{v}\times\bm B(\bm{r})],
 \label{eq:motion}
\end{equation}
where $\bm p$ is the particle momentum, $\bm v$ is the particle velocity, 
$q$ is the particle charge, $\bm E$ and $\bm B$ are the convection electric
field and total magnetic field, respectively, $\bm r$ is the particle position 
in the shock reference frame, and $t$ is time.
The convection electric field satisfies $\bm E=-\bm U\times\bm B$, 
where $\bm U$ is the bulk fluid velocity in the shock frame. We consider
a planar shock located at $z=0$ for simplicity. The shock geometry 
is illustrated in Figure \ref{fig:shockdbB0}(a) \citep[e.g.,][]{KongAQin2020,QinEA2023}.  
The bulk fluid flows in the positive $z$-direction with the upstream speed $U_1$
and downstream speed $U_2$ in the shock frame. The fluid speed
in the shock transition is assumed to be in the form of  
\begin{equation}
U(z)=\frac{U_1}{2s}\left\{\left(s+1\right)+\left(s-1\right)\tanh
\left[\tan\left(-\frac{\pi z}{L_{\text{th}}}\right)\right]\right\},
\end{equation}
where $L_{\text{th}}$ is the shock thickness, and $s$ is the shock compression ratio.
The magnetic field is given by
\begin{equation}
\bm B(x,y,z)=\bm B_0(z)+\bm b(x,y,z),
\end{equation}
where $\bm B_0$ is the background field located in the $x$-$z$ plane, and $\bm b$ is
a zero-mean turbulent ﬁeld. The turbulent field $\bm b$ is composed of slab and
2D components \citep{MatthaeusEA1990,MaceEA2000,QinEA2002a,QinEA2002b}, and follows 
a Kolmogorov spectrum with a spectral index $-5/3$ at high wavenumbers. The magnetic 
field $\bm B_2$ and bulk fluid velocity $\bm U_2$ downstream of the shock are obtained 
from those values upstream of the shock via the MHD Rankine-Hugoniot (RH) conditions.

We assume that the correlation length of slab turbulence, $\lambda_{\text{slab}}$, 
is proportional to the distance to the Sun. Therefore, the slab correlation length 
is assumed to be $2.1\times10^8$ m at 0.07 au and $2.1\times10^9$ m at 0.7 au, 
scaled from the commonly adopted value of 0.02 au at 1 au, as listed in Table \ref{tab:shockpara}.
The turbulence is generated based on the fast Fourier transform (FFT) in a 
periodic box with a size of $50\lambda_{\text{slab}}$ for the slab component
and sizes $10\lambda_{\text{slab}}\times10\lambda_{\text{slab}}$ for the 2D component.
The 2D correlation length is set to be $\lambda_{\text{slab}}/2.6$ and the ratio
of slab to 2D magnetic energy densities in the turbulence is set to be 
$E_{\text{slab}}:E_{\text{2D}} = 20:80$.
The turbulence level, $(b/B_0)^2$, as a ‌crucial parameter, directly determines the
acceleration process of particles. In the previous works, both an inverse proportional 
function that decreases with distance from the shock front and a constant profile 
have been used \citep{KongEA2017,QinEA2018,KongAQin2020}. 
In this work, we describe the amplitude of magnetic field fluctuations $b/B_0$ 
using an exponential decay function
\begin{equation}
b/B_0=C\big\vert[\exp(a_0z)+a_1]\big\vert,
\label{equ:bB0}
\end{equation} 
where for simplicity the parameters are specified as $C=0.5$, $a_1=0.02$,
and $a_0=5\times10^{-5}$ km$^{-1}$ for $z<0$ 
and $a_0=-5\times10^{-5}$ km$^{-1}$ for $z>0$. 
As plotted in Figure \ref{fig:shockdbB0}(b), the fluctuation amplitude 
$b/B_0$ shows a sharp decline near the shock and decays to a very low value 
of 0.01 far away from the shock.

The particle fluxes observed by PSP at the shock crossing on 2022 September 5 
and by SolO on 2022 September 6, in the energy channels listed in Table \ref{tab:Echannels}, 
are simulated using a backward-in-time method assuming an initial distribution. 
We assume an initial kappa distribution constructed from the upstream density, 
temperature and solar wind momentum in the spacecraft frame \citep{Neergaard-ParkerEA2014,KongEA2017}.  
In the simulations, a total of $10^6$ test protons for the PSP and 
SolO events, respectively, each with a target energy $E_i$, were 
isotropically released at the location $z=L_{\text{th}}$. 
The trajectory of each particle is traced via Equation \ref{eq:motion} using
the fourth-order Runge–Kutta method with an adaptive step that maintains 
a normalized accuracy of $10^{-9}$ within the simulation time. 
The distribution of particles accelerated by the shock, $f(E_i)$, 
is obtained as
\begin{equation}
f(E_i)=\frac{1}{N_i}\sum_{j=1}^{N_i}f_0(E_{ij}),
\end{equation}   
where $E_i$ represents the central energy of the $i$th channel, $N_i$ is 
the number of test particles in each energy channel, $f_0$ is the initial 
distribution, and $E_{ij}$ is the energy of the $j$th particle when it is 
traced back to the initial time. 
Thus, the corresponding energy spectrum is given by
\begin{equation}
j(E_i)=f(E_i)p_i^2,
\end{equation}
where $p_i$ is the particle momentum.
The initial distribution $f_0$ is assumed to be
\begin{equation}
f_0(\boldsymbol p)=C\left[1+\frac{(\boldsymbol p-\boldsymbol p_0)^2}{(2\kappa-3)mkT}\right]^{-\kappa-1},
\label{equ:f0}
\end{equation}
where $C$ is the normalization constant given by
\begin{equation}
C=n\left[\frac{1}{(2\kappa-3)\pi mkT}\right]^{3/2}\frac{\Gamma(\kappa+1)}{\Gamma(\kappa-1/2)},
\end{equation}
$\bm p$ is the particle momentum, $\bm p_0$ is the solar wind momentum 
with a speed $U_{\text{sw}}$, $n$ is the number density, $T$ is the 
temperature of the solar wind proton population, $m$ is proton mass, 
and $k$ is the Boltzmann constant.
Note that a few particles, which have a velocity lower than 
the background bulk flow speed during the simulations, are abandoned. 
In addition, the wave excited by the accelerated particles
is not considered in this work.

\begin{deluxetable*}{ll}
\tablecaption{Energy Channels\label{tab:Echannels}}
\tablehead{
\colhead{Instrument} & \colhead{$E_i$ (MeV)}}
\startdata
PSP/EPI-Lo    & 0.0697\;0.0927\;0.142\;0.195\;0.280\;0.422\;0.659 \;1.040 \;1.66 \;2.66 \;4.27\;5.86\;8.06\\
SolO/EPD   & 0.0495\;0.0729\;0.103\;0.142\;0.236\;0.322\;0.526 \;0.804 \;1.03 \;1.46 \;2.05\;2.92\;4.10\;5.83
\enddata
\end{deluxetable*}

\section{Results} \label{sec4}

\subsection{Comparisons of PSP-SolO Observations and Simulations}

\begin{figure*}[ht!]
 \centering
\includegraphics[scale=0.6]{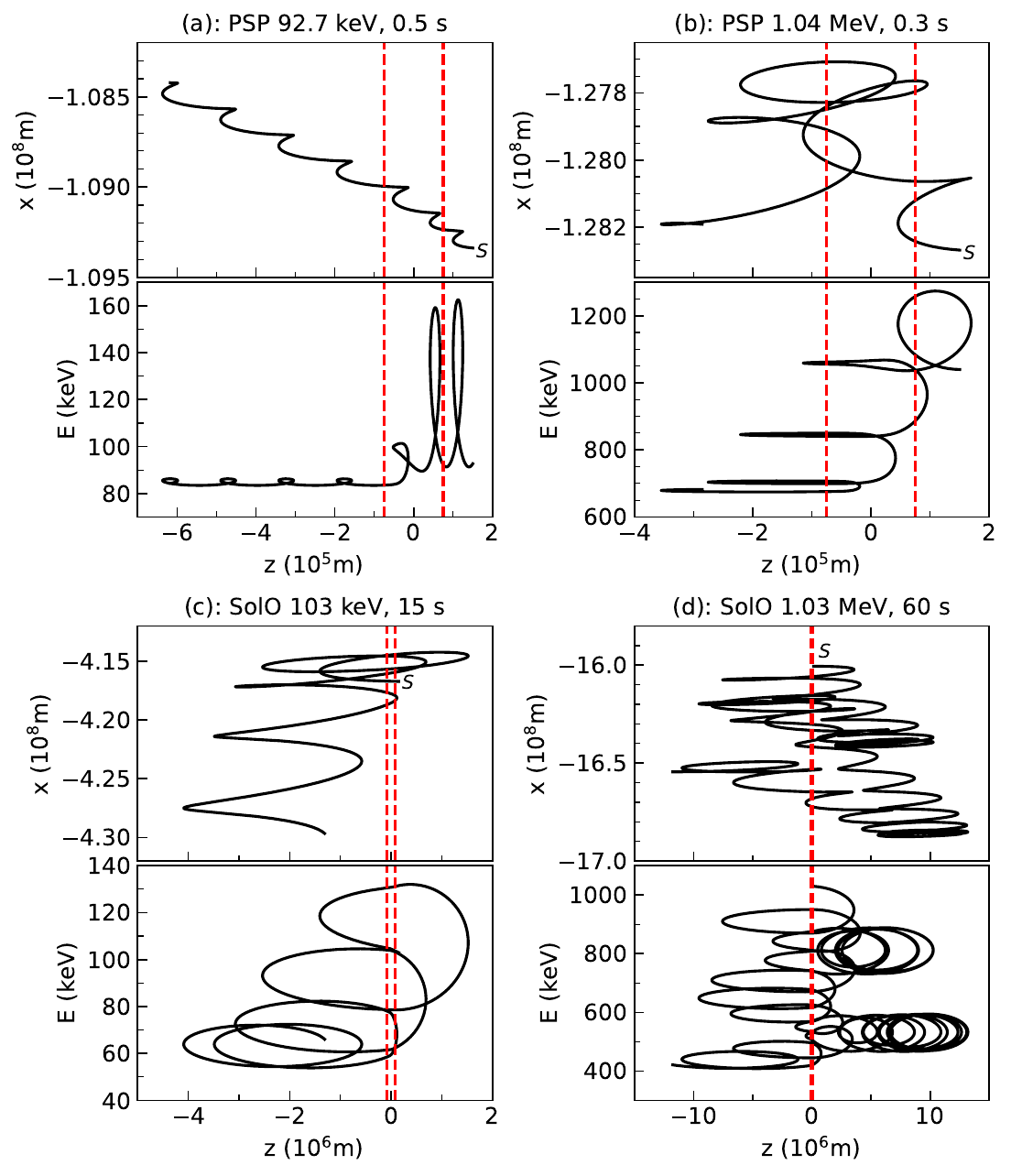}
\caption{Sample trajectories of four test particles in the energy channels centered 
at 92.7 keV (a) and 1.04 MeV (b) at PSP, and 103 keV (c) and 1.03 MeV (d) at SolO.
Upper panel: the $x$-component vs. $z$-component of the particle position
in the shock frame, where the region between the two vertical dashed lines 
indicates the shock transition. Lower panel: particle energy $E$ in the 
spacecraft frame vs. the $z$-component. 
\label{fig:orbit}}
\end{figure*}

Figure \ref{fig:orbit}(a--d) shows sample proton trajectories of four test particles 
in the energy channels centered at 92.7 keV and 1.04 MeV at PSP, and at
103 keV and 1.03 MeV at SolO. In each panel, the upper panel plots the $x$-component 
of particle position against the $z$-component in the shock frame, while the lower
panel shows the particle energy $E$ in the spacecraft frame as a function of the $z$-component. 
Note that the region between the two vertical dashed lines marks the shock transition.
As shown in the lower panel of each panel, the particle is released at point $S$ in 
the downstream region. The particle energy decreases during the simulation,
as we employ the backward-in-time method.
The particle shown in Figure \ref{fig:orbit}(a) undergoes a single crossing 
of the shock transition and then escapes to the upstream region within a total time 
of 0.5 s, with its energy decreasing from the target energy of 92.7 keV to $\sim$84 keV. 
Figure \ref{fig:orbit}(b) shows that within 0.3 s, the particle has a notable 
energy decrease from $\sim$1 MeV to $\sim$682 keV through three crossings of the shock front.
In Figure \ref{fig:orbit}(c), the particle undergoes 7 crossings of the 
shock transition within 15 s, and its energy decreases from 103 keV to $\sim$ 66 keV. 
Figure \ref{fig:orbit}(d) shows a particle trajectory within 1 minute, during which
the particle energy falls from $\sim$1 MeV to $\sim$420 keV through 17 crossings 
of the shock transition. 
In the $E$--$z$ plot for the 103 keV and 1.03 MeV channels in 
Figures \ref{fig:orbit}(c) and (d), the particle trajectories exhibit repeated 
back-and-forth crossings of the shock transition. This behavior arises from
the larger particle gyroradius in the weaker ambient magnetic field at 0.7 au, 
which enables particles to easily traverse the shock transition. The shock drift 
acceleration plays a dominant role for particles with large gyroradii‌.

\begin{figure*}[ht!]
 \centering
\includegraphics[scale=0.6]{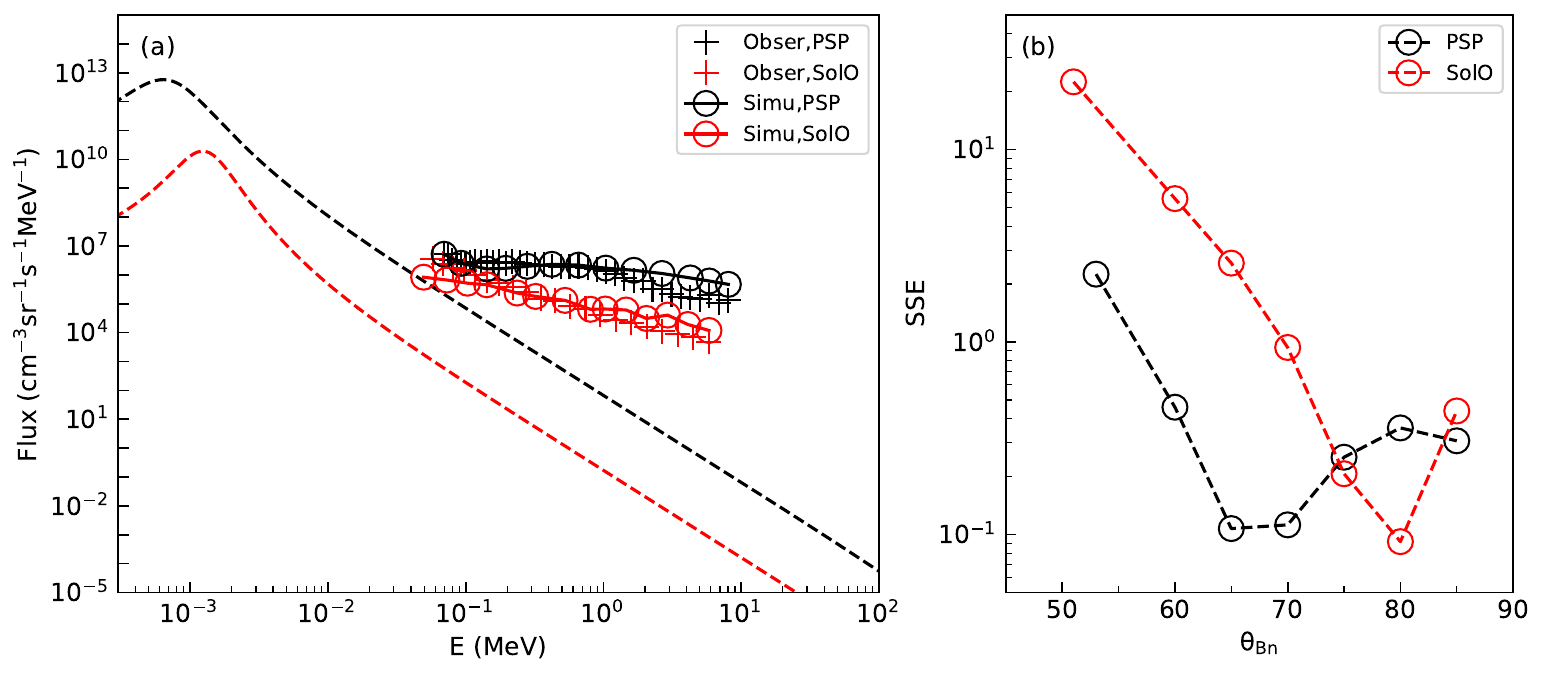}
\caption{Left panel: 
Proton energy spectra downstream of the shock at $z=L_{\text{th}}$ 
from the simulations for the PSP-observed 2022 September 5 shock 
(black circles) and SolO-observed 2022 September 6 shock (red circles) are shown.
The assumed initial $\kappa=3$ kappa distributions are plotted with
the corresponding colored dashed curves.
Also shown are energy spectra from the observations just at the shock arrival
for the 2022 September 5 shock (black crosses), measured by PSP/IS$\Theta$IS/EPI-Lo
instrument, and for the 2022 September 6 shock (red crosses), measured by SolO/EPD/EPT instrument.
Right panel: Variations of the Sum of Squares for Error (SSE) with shock angle for
PSP (black circles) and SolO (red circles) events. SSE is calculated as the sum of 
squares of the differences between the simulated and observed fluxes for 
all energy channels listed in Table \ref{tab:Echannels}, excluding the 
49.5 keV channel for the SolO-observed shock event due to missing observed fluxes.
\label{fig:fluxE}}
\end{figure*}

Figure \ref{fig:fluxE}(a) shows the proton energy spectra at $z=L_{\text{th}}$ 
downstream of the shock for the simulations (circles) and at the shock
arrival for the observations (crosses). A kappa initial distribution with $\kappa=3$
is constructed using Equation \ref{equ:f0} for the PSP-observed shock at 0.07 au 
on 2022 September 5 (black dashed line) and SolO-observed shock at 0.7 au 
on 2022 September 6 (red dashed line).
The PSP-observed shock at 0.07 au and SolO-observed shock at 0.7 au 
are denoted as Event 1 and Event 2, respectively, in the following discussion. 
For both Event 1 and Event 2, the energy spectra obtained from the simulations
(circles) are in good agreement with the observational spectra (crosses),
‌though with some discrepancies. In the PSP-observed Event 1, the simulated fluxes 
are slightly higher than the observed fluxes at energies above $\sim$1 MeV. In the 
SolO-observed Event 2, the simulations yields a lower flux below $\sim$0.2 MeV 
and a higher flux above $\sim$1 MeV compared with the observational flux.
Note that shock angles of 70$^\circ$ for Event 1 and 80$^\circ$ for Event 2 
are adopted in the simulations, as the simulations failed to reproduce 
the observed spectra using the angles listed in Table \ref{tab:shockpara}.
This choice is justified by a quantitative goodness-of-fit analysis based 
on the Sum of Squares for Error (SSE), as detailed below.

In Figure \ref{fig:fluxE}(b), we plot the variations of the 
Sum of Squares for Error (SSE) as a function of the shock angle ‌at 
5$^\circ$ intervals for the PSP-observed Event 1 (black circles) 
and SolO-observed Event 2 (red circles). The SSE is defined as 
\begin{equation}
SSE=\sum_i[\log_{10}F_{\text{simu},i}-\log_{10}F_{\text{obser},i}]^2,
\end{equation}
where $F_{\text{simu},i}$ and $F_{\text{obser},i}$ are the simulated and 
observed fluxes in the $i$-th energy channel, respectively. All energy channels 
listed in Table \ref{tab:Echannels} are included in this calculation, 
excluding the 49.5 keV channel for Event 2 due to the absence of observational flux data. 
In the simulations for each shock angle, an acceleration time of
1 hour was set to ensure that the resulting accelerated energy spectrum
exhibits no notable time-dependent variation.
As shown in Figure \ref{fig:fluxE}(b), the minimum SSE is achieved at 
a shock angle of 80$^\circ$ for the SolO-observed Event 2, while the SSE values at 65$^\circ$ and 
70$^\circ$ for the PSP-observed Event 1 are close to the minimum. Accordingly, shock angles of 
$70^\circ$ and $80^\circ$ are selected for Event 1 and Event 2, respectively, 
in the simulations. Under these two quasi-perpendicular shock geometries, 
we find that a 10-minute acceleration duration is sufficiently long for the 
shock to energize particles until a steady-state energy spectrum is established.

\begin{figure*}[]           
 \centering
\includegraphics[scale=0.6]{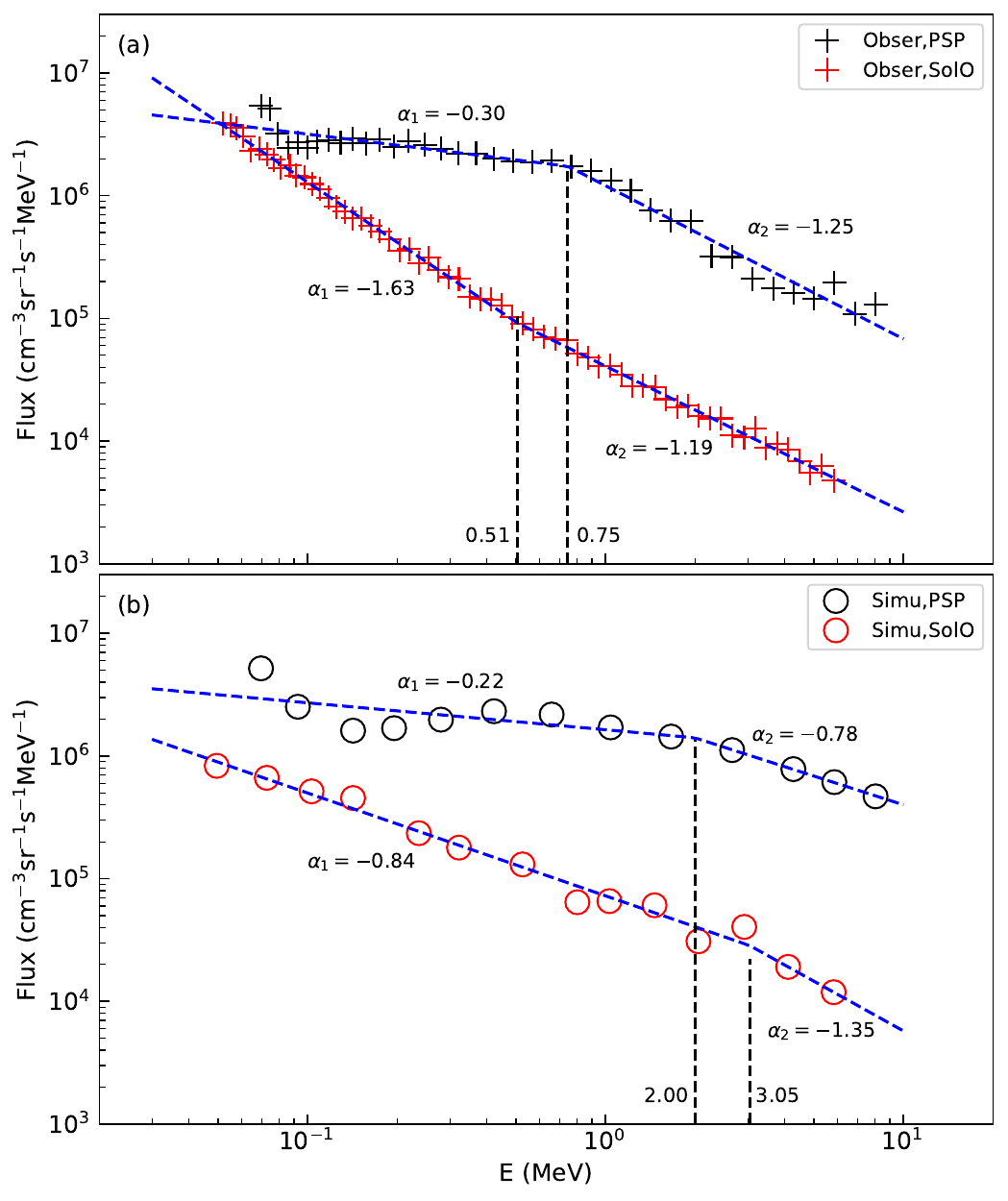}
\caption{Double power-law fittings of the energy spectra for
the observations (Panel (a)) and simulations (Panel (b)).
The vertical dashed lines mark the spectral breaks of the double 
power-law fits, and both the slopes and break energies 
in the log-log plot are labeled.
\label{fig:fluxEfit}}
\end{figure*}

Figure \ref{fig:fluxEfit} shows the double power-law fits for both observed
and simulated energy spectra. In Figure \ref{fig:fluxEfit}(a), the PSP-observed
spectrum (black crosses) is softer above the break energy of $\sim$0.75 MeV, 
while the SolO-observed spectrum (red crosses) is harder above $\sim$0.51 MeV. 
As mentioned in Section \ref{sec2}, there is an inverse dispersion relation for 
particles above $\sim$1 MeV observed by PSP. 
A longer acceleration time is required for high-energy particles to be effectively 
accelerated and subsequently released by the shock. For PSP observations located in
the near-Sun region, high-energy particles above 1 MeV, compared with particles below 1 MeV,
cannot be confined around the shock long enough to complete sufficient acceleration.
This directly contributes to a relatively softer energy spectrum in the high-energy range. 
On the other hand, a large population of seed particles can be efficiently accelerated 
in the low-energy channels, whereas the source of the seed particles 
for the high-energy channels is intrinsically insufficient. 
As shown in Figure \ref{fig:fluxtsolo}, the particle fluxes above 526 keV 
observed by SolO exhibit no distinct peaks at the shock passage, indicating
that the local shock has negligible contribution to the acceleration of 
particles beyond 526 keV. 
A fraction of high-energy particles pre-trapped in the interplanetary 
magnetic field are advected to SolO at the shock arrival, which
yields a harder energy spectrum in the higher-energy channels.
The simulated energy spectrum (Figure \ref{fig:fluxEfit}(b)) at PSP is harder
than the observed one (Figure \ref{fig:fluxEfit}(a)), with a higher break
energy of up to 2 MeV. At SolO, the simulations show a harder spectrum 
than the observations below a break energy of 3.05 MeV, which also exceeds 
the fitted observational value. 
The higher break energies for both PSP and SolO in the simulation 
are likely due to the quasi-perpendicular shock angles of $70^\circ$ and 
$80^\circ$ used in the simulations, which contribute to more efficient acceleration by the shock. 

The shock angles used in the simulations ($70^\circ$ for the PSP-observed Event 1 and $80^\circ$
for the SolO-observed Event 2) are larger than the in-situ values ($51^\circ$ and $53^\circ$) 
measured at the shock arrival of PSP and SolO. This discrepancy can be physically explained as follows.
Both the small-scale local shock ripples ubiquitous in the
inner heliospheric turbulent environment and large-scale upstream
magnetic field fluctuations conspire to form localized quasi-perpendicular
geometries that are probably not directly traversed by the spacecraft. 
The sampled high-energy particles, energized at those off-spacecraft 
quasi-perpendicular shock regions, subsequently propagate along the 
connected magnetic field lines to the spacecraft.
We note that both the observed and simulated spectral indices 
(marked in Figure \ref{fig:fluxEfit}) are higher than the standard DSA 
prediction, which follows the relation $j(E)\propto E^{-(s+2)/2(s-1)}$ and 
yields values of $-1.65$ for the PSP-observed shock and $-2.17$ for the SolO-observed shock.
This result indicates the energy spectra for both observations and simulations
are harder than the DSA theoretical prediction. 
The reason is probably non-local particle acceleration at large-angle 
quasi-perpendicular shocks, the efficient injection of pre-accelerated 
suprathermal particles, and the finite shock lifetime effect.
Particles accelerated at quasi-perpendicular shocks in regions away from the 
immediate shock front propagate to the observation site, significantly enhancing 
the flux of high-energy particles. Meanwhile, pre-existing
suprathermal seed particles can directly enter the DSA process, increasing the 
fraction of high energy particles without being limited by the low-energy injection threshold.
The finite shock lifetime enables preferential escape of low-energy particles, 
while high-energy populations remain effectively confined by magnetic turbulence 
in the vicinity of the shock, naturally leading to the observed hardened energy spectrum.
The initial $\kappa=3$ distribution used in our simulations, which contains a larger
population of suprathermal particles in the high-energy tail than the Maxwellian distribution,
readily yields a harder spectrum than predicted by standard DSA theory.

\begin{figure*}[ht!]
 \centering
\includegraphics[scale=0.6]{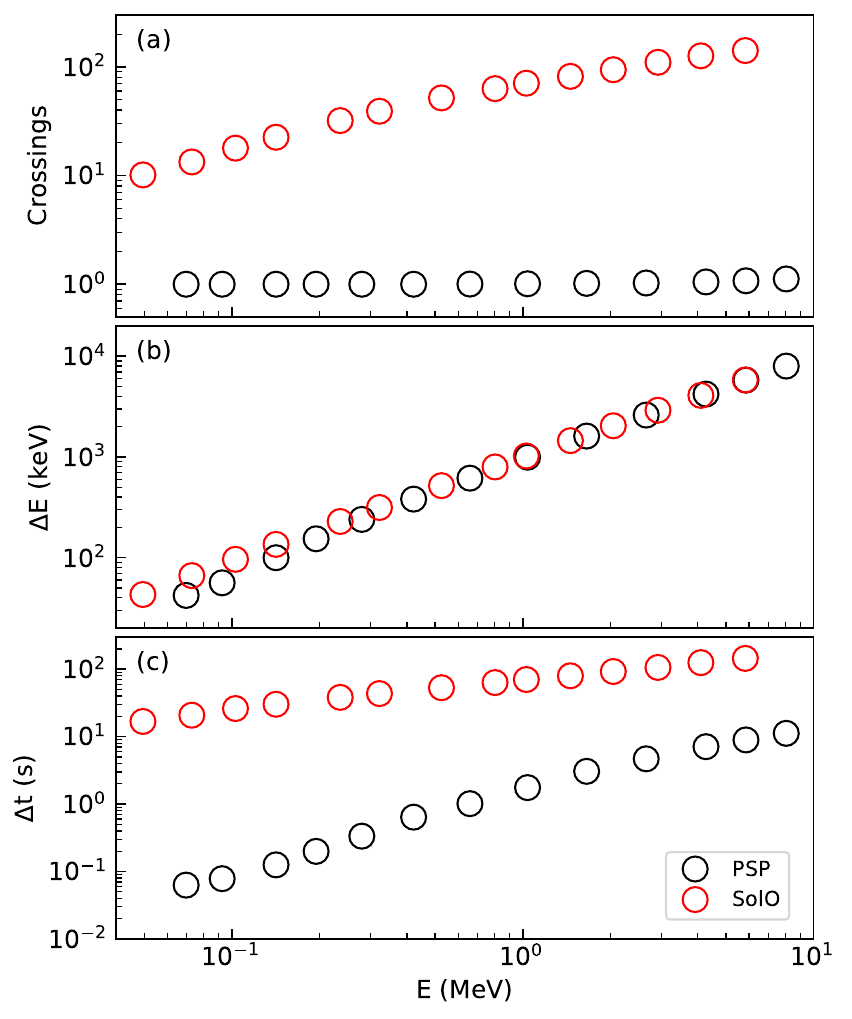}
\caption{Average crossings of the shock transition (a), average 
energy gain $\Delta E$ (b), cumulative‌ residence time‌ $\Delta t$ (c) 
within one gyroradius of the shock front‌ in each energy channel 
considered in the simulations.
\label{fig:dEdt}}
\end{figure*}

In Figure \ref{fig:dEdt}, we present average crossings of the shock transition 
(panel a), average particle energy gain $\Delta E$ (panel b), and cumulative residence time
$\Delta t$ of particles within one gyroradius of the shock front (panel c) for each energy channel
for Event 1 observed by PSP (black circles) and Event 2 observed by SolO (red circles) 
in the simulations by using the initial $\kappa$-distribution as the weight. 
As shown in the energy channels in Figure \ref{fig:dEdt}(a), there are more 
crossings of the shock transition in the SolO-observed Event 2 compared with 
the PSP-observed Event 1‌, where only one shock crossing occurs. The number of 
shock crossings increases with particle energy in Event 2, ranging from 10 to over 100. 
In Figure \ref{fig:dEdt}(b), the average energy gain, $\Delta E$, increases with 
particle energy $E$ for both the PSP-observed Event 1 and SolO-observed Event 2.
Interestingly, the energy gain $\Delta E$ in Event 1 is ‌comparable to‌ that in Event 2.
Although the two events studied show different shock 
crossings, with Event 1 containing very few high-gain crossings and 
Event 2 featuring multiple low-gain crossings, the $\Delta E$--$E$ 
relation in both events follows the same linear trend.
This result can be attributed to the significant contribution of the 
suprathermal tail in the initial kappa distribution, which naturally 
suppresses the energization discrepancy for distinct shock parameters.
Figure \ref{fig:dEdt}(c) shows the cumulative residence time $\Delta t$ of 
particles within one gyroradius of the shock front, for both the PSP-observed 
Event 1 and SolO-observed Event 2. In Event 1, $\Delta t$ increases gradually 
from $\sim$0.06 s to $\sim$11 s, while in Event 2 it remains at a high level 
spanning $\sim$10 s to $\sim$150 s. This result further confirms that the 
acceleration processes in these two events are fundamentally distinct.

Particles in the PSP-observed Event 1 increase their energy within a single
shock crossing, which naturally yields a high escape probability and thus 
produces a lower high-energy cutoff (break energy) of $\sim$2.0 MeV in 
the energy spectrum, as marked in Figure \ref{fig:fluxEfit}(b). 
In contrast, particles in the SolO-observed Event 2 are confined near the
shock front long enough to gain energy via turbulence scattering during
multiple shock crossings. This effect produces a higher break energy of 
$\sim$3.0 MeV in the energy spectrum, as also indicated in Figure \ref{fig:fluxEfit}(b).

\subsection{Dependences of $\Delta E$ and $\Delta t$ on $\theta_{\text{Bn}}$, $s$ and $M_{\text A1}$}

We further investigate the dependences of average energy gain $\Delta E$
and cumulative residence time $\Delta t$ within one gyroradius of the shock front 
on the shock angle $\theta_{\text{Bn}}$, compression ratio $s$, and 
upstream Alfv$\acute{\text{e}}$n Mach number $M_{\text A1}$ 
at two distinct locations of 0.07 au and 0.7 au. A range of values 
for $\theta_{\text{Bn}}$, $s$, and $M_{\text A1}$, used in the simulations, 
are listed in Table \ref{tab:thetaMa1s}. 
Assume that the shock speed satisfies the relation
\begin{equation}
V_{\text{sh}}=M_{\text A1}V_{\text A1}+U_{\text{sw}},
\label{equ:VshMa1}
\end{equation} 
where $V_{\text A1}$ is the upstream Alfvén speed.
In Equation \ref{equ:VshMa1}, both the shock speed and solar wind speed 
are defined in the spacecraft reference frame, with the requirement that 
the shock speed exceeds the solar wind speed.
The shock speed $V_{\text{sh}}$ varies with Mach number $M_{\text A1}$, 
provided that both the Alfvén and solar wind speeds are fixed. 
The corresponding shock speeds, $V_{\text{sh}}$, computed over a range of 
Mach numbers when using the Alfvén and solar wind speeds in Event 1 
(PSP at 0.07 au) and Event 2 (SolO at 0.7 au), are also listed in Table \ref{tab:thetaMa1s}.
In examining how $\Delta E$ and $\Delta t$ vary with $\theta_{\text{Bn}}$, $s$, 
or $M_{\text A1}$, we fix all other shock parameters using values from Event 1 
at 0.07 au and Event 2 at 0.7 au, respectively, yielding distinct simulation results.

\begin{deluxetable*}{ll}
\tablecaption{Parameter Values for Various Cases\label{tab:thetaMa1s}}
\tablehead{
\colhead{Parameters} & \colhead{Values}}
\startdata
$\theta_{\text{Bn}}$  & $3^\circ$\; $15^\circ$\; $30^\circ$\; $45^\circ$\; $60^\circ$\; $75^\circ$\; $87^\circ$\\
$s$  & 1.2\; 1.5\; 2.0\; 2.5\; 3.0\; 3.5\; 3.9\\
$M_{\text A1}$  & 2.0\; 3.0\; 4.0\; 5.0\; 6.0\\
($V_{\text{sh}}$ km s$^{-1}$ PSP/SolO) & 1100/812\; 1480/978\; 1860/1144\; 2240/1310\; 2620/1476\\
\enddata
\end{deluxetable*}

\begin{figure*}[ht!]
 \centering
\includegraphics[scale=0.6]{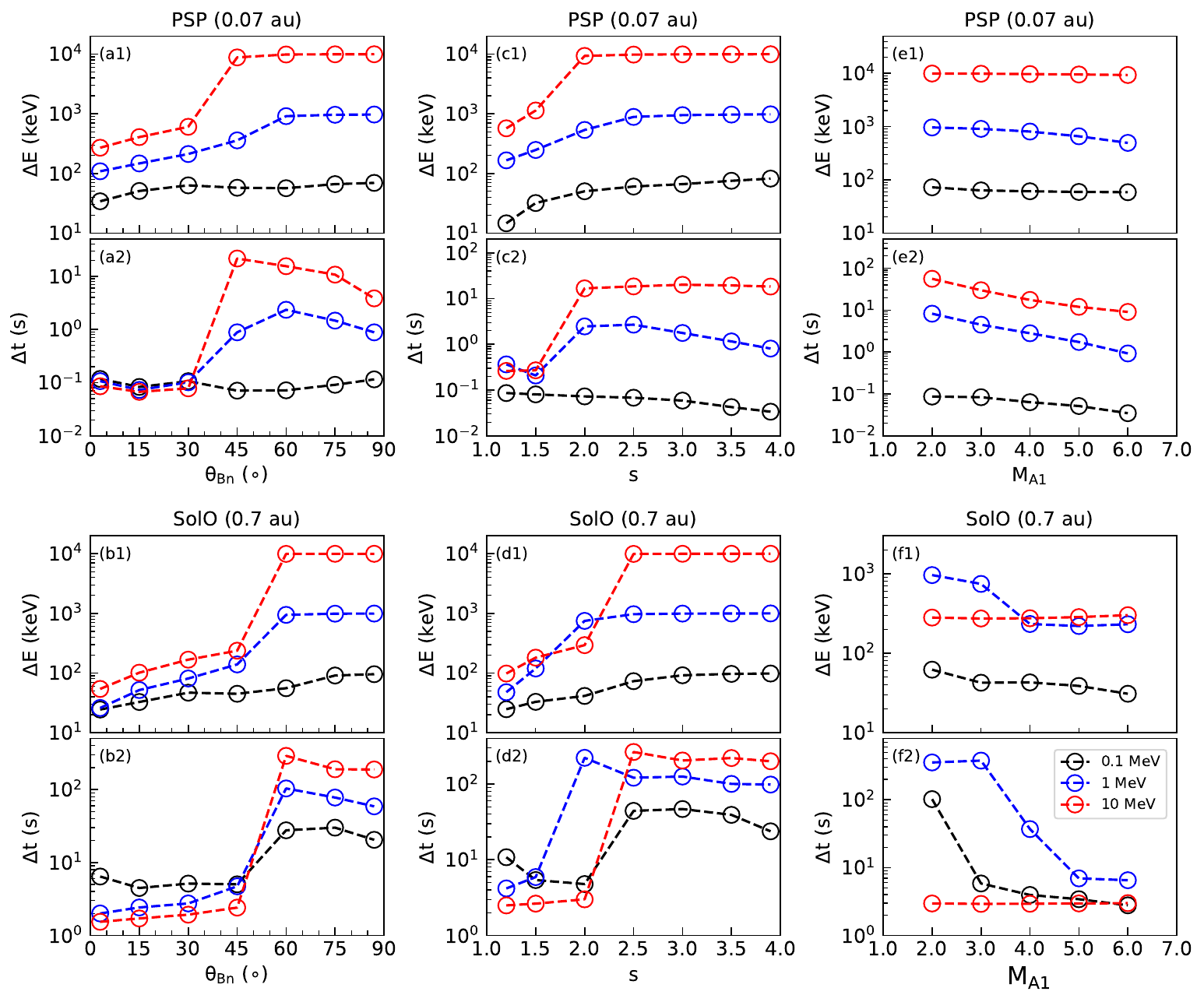}
\caption{Dependencies of the average energy gain $\Delta E$ and 
cumulative‌ residence time‌ $\Delta t$ (within one gyroradius of the shock front‌)
on three shock parameters. Panels (a1)--(b2) show variations with the shock angle 
$\theta_{\text{Bn}}$, (c1)--(d2) with the shock compression ratio $s$, and (e1)--(f2)
with the upstream Alfvén Mach number $M_{\text{A1}}$.
Two independent sets of results are presented, using PSP (0.07 au) and 
SolO (0.7 au) observational parameters, respectively. The 0.1 MeV, 1 MeV 
and 10 MeV energy channels are denoted by black, blue and red circles, respectively.
\label{fig:dEdtpara}}
\end{figure*}

Figure \ref{fig:dEdtpara} presents two simulation results obtained 
using PSP and SolO in-situ observational parameters, respectively.
It shows the average energy gain $\Delta E$ and cumulative‌ residence 
time‌ $\Delta t$ within one gyroradius of the shock front‌, 
calculated at various shock angles ranging from $3^\circ$ to $87^\circ$, 
shock compression ratio $s$ from 1.2 to 3.9, and upstream Alfvén Mach number 
$M_{\text A1}$ from 2.0 to 6.0, for particles with an energy of 0.1 MeV (black circles), 
1 MeV (blue circles) and 10 MeV (red circles).
In Figure \ref{fig:dEdtpara}(a1) and (b1), the particle energy gain 
$\Delta E$ increases monotonically with the shock angle $\theta_{\text{Bn}}$ 
before reaching its peak for all three energy channels, using the PSP 
(0.07 au) parameter set for panel (a1) and SolO (0.7 au) set for panel (b1).
The energy gain of 0.1 MeV particles shows the smallest growth, 
remaining within the range of $10^1$--$10^2$ keV and barely rising further 
at large angles at both 0.07 au and 0.7 au. The energy gain of 1 MeV 
particles at 0.07 au gradually increases from $\sim$100 keV
up to $\sim$1000 keV, while at 0.7 au it rises from $\sim$20 keV
to the same upper limit. The energy gain $\Delta E$ of 10 MeV particles 
exhibits the most significant growth, rising rapidly to $\sim10^4$ keV 
when $\theta_{\text{Bn}}$ exceeds $30^\circ$ (at 0.07 au) and $45^\circ$ (at 0.7 au), respectively. 
The cumulative‌ residence time‌ $\Delta t$ for the PSP (0.07 au) parameter set 
in Figure \ref{fig:dEdtpara}(a2) almost completely overlaps across all 
three energy channels at $\theta_{\text{Bn}}<30^\circ$, maintaining an 
extremely low level around $\sim$0.1 s. Once $\theta_{\text{Bn}}$ exceeds $30^\circ$, 
$\Delta t$ for the 1 MeV and 10 MeV particles increases rapidly, 
peaking at approximately $60^\circ$ and $45^\circ$ respectively, 
before gradually declining at larger $\theta_{\text{Bn}}$. The overall 
trend of $\Delta t$ for the SolO parameters at 0.7 au in Figure \ref{fig:dEdtpara}(b2) 
is consistent with that for PSP parameters at 0.07 au, with 
the exception of 0.1 MeV particles. However, the absolute magnitude 
of $\Delta t$ is 1--2 orders of magnitude higher at 0.7 au than at 0.07 au, 
and the peak of $\Delta t$--$\theta_{\text{Bn}}$ distribution at 0.7 au 
is located near $\theta_{\text{Bn}}\sim60^\circ$. The increase of $\Delta t$ 
in the 0.1 MeV particles is far more notable at 0.7 au than at 0.07 au. 
 
As shown in Figure \ref{fig:dEdtpara}(c1) and (d1), the particle energy gain $\Delta E$ 
for all three energy channels increases with the compression ratio $s$, and then
exhibits no noticeable variation for $s>2.5$. For the 0.1 MeV particles, $\Delta E$ 
shows the weakest variation with $s$ at both 0.07 au and 0.7 au; 
for the 10 MeV particles, $\Delta E$ exhibits a prominent increase
for $s>1.5$ at 0.07 au and for $s>2.0$ at 0.7 au. 
At 0.07 au (Figure \ref{fig:dEdtpara}(c2)), the cumulative residence time 
$\Delta t$ for the 0.1 MeV particles decreases slowly with increasing compression 
ratio $s$. For the 1 MeV particles, $\Delta t$ peaks at $s=2.5$ and then decreases 
gradually with the compression ratio, while for the 10 MeV particles, 
$\Delta t$ exhibits a large increase when $s$ varies from 1.5 to 2.0, after which
it shows no pronounced variation.
In contrast, at 0.7 au (Figure \ref{fig:dEdtpara}(d2)), $\Delta t$ for all 
three energy channels shows a sharp increase of 1–2 orders of magnitude, 
followed by a gradual decrease after the peak.

In Figure \ref{fig:dEdtpara}(e1) at 0.07 au, the energy gain $\Delta E$ 
shows no obvious variation with the upstream Alfvén Mach number $M_{\text A1}$ 
for all three energy channels. In contrast, in Figure \ref{fig:dEdtpara}(f1) at 0.7 au, 
$\Delta E$ for the 1 MeV particles decreases markedly from $\sim1000$ keV to $\sim200$ keV
as $M_{\text A1}$ increases from 2.0 to 4.0. At the same heliocentric distance, 
$\Delta E$ for the 0.1 MeV particles decreases slowly, while for the 10 MeV particles, 
$\Delta E$ exhibits no significant variation with $M_{\text A1}$.
Compared to the gradual decrease of the cumulative residence time $\Delta t$ with $M_{\text A1}$
for particles at all three energy channels at 0.07 au in Figure \ref{fig:dEdtpara}(e2), 
$\Delta t$ at 0.7 au in Figure \ref{fig:dEdtpara}(f2) shows a distinct trend with $M_{\text A1}$:
$\Delta t$ for the 0.1 MeV particles drops from 100 s to less than 10 s, 
$\Delta t$ for the 1 MeV particles falls from over 300 s to less than 10 s, 
while $\Delta t$ for 10 MeV particles remains nearly constant at $\sim$3 s.

The average energy gain $\Delta E$ arises from the combined effects of shock
drift acceleration (SDA) and first-order Fermi acceleration (FFA). 
The cumulative‌ residence time $\Delta t$ within one gyroradius of the shock front
directly reflects the contribution of SDA.
At 0.07 au, $\Delta t$ peaks at $\theta_{\text{Bn}}=60^\circ$ for the 1 MeV particles 
and at $\theta_{\text{Bn}}=45^\circ$ for the 10 MeV particles.
At 0.7 au, by contrast, $\Delta t$ peaks consistently at $\theta_{\text{Bn}}=60^\circ$
for all three energy channels. Notably, $\Delta t$ for the 10 MeV particles at 
both heliocentric distances shows the most pronounced enhancement, increasing 
by more than 2 orders of magnitude. The simulation results presented above 
demonstrate that the contribution of drift acceleration is strongly boosted
at oblique and quasi-perpendicular shocks. 
At 0.07 au, $\Delta t$ shows an obvious increase at $s=2.0$ for both the 1 MeV 
and 10 MeV particles. At 0.7 au, however, the rapid increase of $\Delta t$ is 
located at $s=2.0$ for the 1 MeV particles, but at $s=2.5$ for both the 0.1 MeV 
and 10 MeV particles. This indicates that, under moderate compression ratios, 
SDA plays a prominent role for the 1 MeV and 10 MeV particles at 0.07 au, 
and for all three energy channels at 0.7 au. 
At 0.07 au, the SDA contribution weakens gradually with increasing 
$M_{\text A1}$ from 2.0 to 6.0, while $\Delta E$ remains nearly unchanged,
indicating the enhanced FFA contribution. In contrast, at 0.7 au, the SDA 
contribution drops notably with increasing $M_{\text A1}$ for the 0.1 MeV 
and 1 MeV particles, accompanied by a rapid increase in the FFA contribution.
For the 10 MeV particles at 0.7 au, SDA is nearly independent of $M_{\text A1}$.

\section{Summary and Discussion} \label{sec5}

A CME-driven shock on 2022 September 5 was observed by PSP at 0.07 au, 
which is the closest in-situ shock measurement to the Sun to date. 
Subsequently, the same CME-driven interplanetary shock was detected by SolO 
on 2022 September 6 at 0.7 au. PSP observed an inverse dispersion signature 
above $\sim$1 MeV, which probably arises from continuous acceleration at 
the shock front. No distinct peak in the ion flux above $\sim$1 MeV was 
detected by SolO at the shock passage. These observations strongly indicate 
that the measured energy spectra from these two shock events cannot be 
fully accounted for by the local diffusive shock acceleration process alone.  
The observed particle fluxes above $\sim$ 1 MeV at the shock arrival 
likely include contributions from the upstream or pre-accelerated particle populations. 

We numerically investigate the acceleration of protons in the energy ranges of 
0.0697--8.06 MeV measured by PSP/EPI-Lo instrument and 0.0495--5.83 MeV 
measured by SolO/EPD instrument. A backward-in-time method is used to 
calculate the particle flux downstream of the shock for each energy channel, 
assuming an initial kappa distribution with $\kappa=3$.
Our simulations successfully reproduce the observed energy spectra at 
the shock arrival for the 2022 September 5 (PSP at 0.07 au) and 2022 
September 6 (SolO at 0.7 au) events, requiring effective quasi-perpendicular 
shock angles ($70^\circ$ and $80^\circ$) that are substantially larger than 
the corresponding in-situ measured values (53$^\circ$ and 51$^\circ$). 
This discrepancy most likely arises from turbulent shock ripples and upstream 
magnetic fluctuations, which contribute to localized quasi-perpendicular 
regions that are not directly traversed by the two spacecraft.

Both PSP at 0.07 au and SolO at 0.7 au observed a double power-law 
energy spectrum at the shock arrival. However, the 
near-Sun PSP and far-Sun SolO measurements exhibit different spectral hardness. 
The energy spectrum at 0.07 au softens above a break energy of $\sim$0.75 MeV, 
whereas the spectrum at 0.7 au hardens above $\sim$0.51 MeV. This discrepancy 
probably stems from different processes of particle acceleration and transport 
at varying heliocentric distances. The near-Sun high-energy particles above 
1 MeV cannot be confined near the shock front long enough to undergo sufficient 
acceleration, which naturally suppresses the high-energy seed population and 
leads to a softer spectral slope at higher energies at 0.07 au. By contrast, 
pre-trapped high-energy particles in the interplanetary space propagate along 
the connected magnetic field lines to SolO at the shock arrival, 
boosting the high-energy particle flux and forming a harder spectrum 
at higher energies at 0.7 au. Compared with in-situ observations, 
the simulations show harder energy spectra below the break energy. 
The break energies from the simulations are 2.0 MeV for the PSP-observed 
event and 3.05 MeV for SolO-observed event, both of which exceed the 
fitted observational values. This is due to the adopted large 
quasi-perpendicular shock angles of 70$^\circ$ for the PSP event 
and 80$^\circ$ for the SolO event, which enable 
more efficient particle acceleration by shocks.
The spectral indices from both simulations and observations are 
higher than the prediction from standard DSA theory.
This can be attributed to non-local particle acceleration at 
large-angle quasi-perpendicular shocks, the efficient injection of 
pre-accelerated suprathermal particles, the finite shock lifetime 
effect, and the initial $\kappa=3$ distribution with an enhanced 
suprathermal tail relative to the Maxwellian distribution.

Focusing on energy channels centered at 0.1 MeV, 1 MeV, and 10 MeV, 
we further investigate the dependences of the average energy gain $\Delta E$
and cumulative residence time $\Delta t$ within one gyroradius of 
the shock front‌ on the shock angle $\theta_{\text{Bn}}$, compression 
ratio $s$, and upstream Alfvén Mach number $M_{A1}$, using shock parameters
representative of the PSP-observed event at 0.07 au and SolO-observed event at 0.7 au. 
The mean energy gain $\Delta E$ arises from both shock drift 
acceleration (SDA) and first-order Fermi acceleration (FFA), with 
the cumulative residence time $\Delta t$ directly reflecting the SDA contribution.
The SDA contribution is strongly enhanced at oblique and quasi-perpendicular shocks. 
Under the moderate compression ratios, SDA plays a dominant role for 1 MeV and 10 MeV 
particles at 0.07 au, as well as for all three energy channels at 0.7 au.
With increasing upstream Alfvén Mach number, SDA weakens and FFA becomes dominant 
at 0.07 au; at 0.7 au, SDA is suppressed for 0.1 MeV and 1 MeV particles, 
while SDA remains independent of the Alfvén Mach number for 10 MeV particles.

%%%%%%%%%%%%%%%%%%%%%%%%%%%%%%%%%%%%%%%%%%%%%%%%%%  fig 1

%%%%%%%%%%%%%%%%%%%%%%%%%%%%%%%%%%%%%%%%%%%%%%%%%% fig 2

%%%%%%%%%%%%%%%%%%%%%%%%%%%%%%%%%%%%%%%%%%%%%%%%%%  fig 3

%%%%%%%%%%%%%%%%%%%%%%%%%%%%%%%%%%%%%%%%%%%%%%%%%%  fig 4

%%%%%%%%%%%%%%%%%%%%%%%%%%%%%%%%%%%%%%%%%%%%%%%%%%  fig 5

%%%%%%%%%%%%%%%%%%%%%%%%%%%%%%%%%%%%%%%%%%%%%%%%%%  fig 6

%%%%%%%%%%%%%%%%%%%%%%%%%%%%%%%%%%%%%%%%%%%%%%%%%%  fig 7

%%%%%%%%%%%%%%%%%%%%%%%%%%%%%%%%%%%%%%%%%%%%%%%%%%  fig 8

%%%%%%%%%%%%%%%%%%%%%%%%%%%%%%%%%%%%%%%%%%%%%%%%%%  fig 9

\clearpage

\begin{acknowledgments}

This work was supported by grants NNSFC 42304173,
NNSFC 42374190, and NNSFC 42474220. The authors thank the support 
of the National Supercomputer Center in Zhengzhou and Tianjin.
The calculations were performed on NSCC-ZZ and TianHe-1 (A).
The Parker Solar Probe and Solar Orbiter data used 
in this work are available via the NASA CDAWeb official 
data archive, hosted by the Space Physics Data Facility (SPDF)
(\url{https://cdaweb.gsfc.nasa.gov/}). 

\end{acknowledgments}

\end{CJK*}
\end{document}